\newif \ifdraft \draftfalse

\documentclass[conference]{IEEEtran}
\IEEEoverridecommandlockouts
\usepackage{graphicx}
\usepackage{balance}  
\usepackage{moreverb}
\usepackage{textcomp}

\usepackage[colorlinks,bookmarksopen,bookmarksnumbered,citecolor=red,urlcolor=red]{hyperref}

\newcommand\BibTeX{{\rmfamily B\kern-.05em \textsc{i\kern-.025em b}\kern-.08em
T\kern-.1667em\lower.7ex\hbox{E}\kern-.125emX}}

\usepackage[all]{nowidow}
\usepackage{epsfig,endnotes}
\usepackage{fancyvrb}

\usepackage{array}

\makeatletter
\newcommand{\thickhline}{%
    \noalign {\ifnum 0=`}\fi \hrule height 1.5pt
    \futurelet \reserved@a \@xhline
}
\newcommand{\mypara}[1]{\textbf{#1}:}

\newcolumntype{"}{@{\hskip\tabcolsep\vrule width 1.5pt\hskip\tabcolsep}}
\makeatother

\usepackage[override]{cmtt}
\usepackage{amsmath}
\usepackage{xspace}
\usepackage{paralist}
\usepackage{color}
\usepackage{listings}
\usepackage{multirow}
\usepackage{highlighter}
\usepackage{wasysym}
\usepackage{rotating}
\usepackage{caption}
  \DeclareCaptionType{copyrightbox}
  \usepackage{subcaption}
\usepackage{balance}
\usepackage{enumitem}
\usepackage{booktabs}

\ifdraft
\newcommand{\todo}[1]{\textcolor{red}{todo: #1}}
\newcommand{\ks}[1]{\textcolor{orange}{ks: #1}}
\newcommand{\tudor}[1]{\textcolor{violet}{tudor: #1}}
\newcommand{\mwh}[1]{\textcolor{blue}{mwh: #1}}

\else
\newcommand{\todo}[1]{}
\newcommand{\tudor}[1]{}
\newcommand{\mwh}[1]{}
\newcommand{\ks}[1]{}
\fi
\newcommand{\code}[1]{\lstinline[columns=fullflexible,basicstyle=\small\sffamily]!#1!}

\newcommand{\tool}{KVolve\xspace}
\newcommand{\redis}{Redis\xspace}
\newcommand{\siqr}[1]{\scriptsize(#1)}
\newcommand{\bench}{\redis-bench\xspace}
\newcommand{\redisFS}{redisfs\xspace}

\newcommand{\tid}{\raise0.5ex\hbox{{\tiny $\sim$}}}

\begin{document}
%
\title{Evolving NoSQL Databases Without Downtime}

\author{\IEEEauthorblockN{Karla Saur}
 \IEEEauthorblockA{Intel Labs\textsuperscript{*}\thanks{*Work performed while at the University of Maryland.}\\
karla.saur@intel.com
}
\and
\IEEEauthorblockN{Tudor Dumitra\c{s} and Michael Hicks}
 \IEEEauthorblockA{University of Maryland, College Park\\
\{tdumitra,mwh\}@umiacs.umd.edu
}
}

\maketitle

\begin{abstract}
  NoSQL  
  databases like Redis, Cassandra, and MongoDB are
  increasingly popular because they are flexible, lightweight, and
  easy to work with. Applications that use these data\-bases
  will evolve over time, sometimes necessitating (or preferring) a change
  to the format or organization of the data. The problem we address in this paper is:
  How can we support the evolution of high-availability applications
  and their NoSQL data \emph{online}, without excessive delays or
  interruptions, even in the presence of backward-incompatible 
  data format changes?  

We present \tool, an extension to the popular \redis NoSQL database, as a
  solution to this problem. \tool permits a developer to
  submit an upgrade specification that defines how to
  transform existing data to the newest version. This transformation
  is applied \emph{lazily} as applications interact with the
  database, thus avoiding long pause times. We demonstrate that \tool is
  expressive enough to support substantial practical updates,
  including format changes to RedisFS, a \redis-backed file system,
  while imposing essentially no overhead in general use and minimal
  pause times during updates.
\end{abstract}


\section{Introduction}
\label{sec:intro}


NoSQL databases, such \redis~\cite{redis}, Cassandra~\cite{cassandra}, and MongoDB~\cite{mongodb},
are increasingly the go-to choice for storing persistent data,
dominating traditional SQL-based database management
systems~\cite{gartner, techrepublic}. NoSQL databases are often
organized as \emph{key-value stores}, in that they provide a simple
key-based lookup and update service (i.e., with ``no SQL''). 
While these databases typically lack a formal schema specification, 
applications attach meaning to the
format of the keys and values stored in the database.
Keys are
typically structured strings, and values store objects represented according to
various formats~\cite{medina2014nosql}, e.g., as Protocol
Buffers (``Protobufs'')~\cite{protobuf}, Thrift~\cite{thrift}, Avro~\cite{avro}, or
JSON~\cite{json} objects.

Database schemas change frequently when applications must support new
features and business needs. For example, multiple schema changes are
applied every week to Google's AdWords database~\cite{F1}. 
Applications that use NoSQL data\-bases also evolve data formats over time, 
and may require
modifying objects to add or delete fields,
splitting objects so they are mapped
to by multiple keys rather than a single key, renaming of keys or
value fields~\cite{sadalage}. When changes are not compatible with the old
version of an application, a straightforward way to deploy them in the field 
would be to shut down the running applications,
migrate each affected object in the database from the old format to
the new format, and then start the new version of the
application. 

High availability applications would prefer to avoid the downtime of
shutdown-and-restart upgrades, but evolving a database on-line is
challenging. Thrift, Protobufs, and Avro provide some support for format changes by
allowing alteration of the data encoding itself or by tracking the
version of an object's ``schema''~\cite{kleppmann, AvroUpd}, but there is still
the task of updating each object in the database (e.g., by iterating
over all of its keys~\cite{redisblog}). For large amounts of data,
this can create an unacceptably long pause. 
As an extreme example, Wikipedia was locked for editing during the upgrade to MediaWiki
1.5, and the schema was converted to the new version in about 22
hours~\cite{Wikipedia2005Downtime}.
Developers could avoid shutting down the application by making
the new format backward-compatible with the old format, but 
this could impose a significant constraint on the future evolution of the application.
It may also be possible to grant applications read-only
access to the old database while the migration takes
place, but applications that have even occasional writes will suffer.

A more general approach to evolve the database online is to migrate data
\emph{lazily}.  When the updated application accesses an object in the old
format, the object is converted to the new format on-the-fly.  Thus, the
long pause due to migrating the data is now
amortized over the updated application's execution, causing slower queries
immediately after the update but no full stoppage. 
Currently, the task of implementing lazy data migration falls on the developer:
applications are rewritten to expect data
in both old and new formats and to migrate from to the new format when
the old one is encountered~\cite{sadalage, servicestack, rethans,
  stack_ov}. 
This approach 
results in code that 
mixes application and format-maintenance logic.
Since there is no
guarantee that all data will ultimately be migrated, the migration
code expands with each format change,
becoming more confusing and harder to maintain.



To address these problems, this paper presents \tool,\footnote{\tool
  stands for \emph{Key-Value store 
  evolution}.} a NoSQL database that provides automatic support for
on-line upgrades using lazy data migration. \tool presents the \emph{logical
view} to applications that \emph{data is at the newest 
version of the format}.
Rather than convert all data at once, keys and values are
converted as they are accessed by the application. Pleasantly, to use
\tool requires almost \emph{no changes
to application code}---they simply indicate the data version they expect
when they connect to the database, and they are permitted to proceed
if their expected version and the logical version
match. When a data upgrade is installed, applications with an
incompatible version must update themselves. They can do this either
by simple stop-and-restart (to the new application version), or they can use
\emph{dynamic software updating}
(DSU)~\cite{DBLP:journals/toplas/HaydenSSHF14,DBLP:conf/middleware/GiuffridaIT14,
Chen:2007:PPL:1248820.1248860, DBLP:conf/oopsla/PinaVH14} or
concurrent application switching (as in parallel AppEngine~\cite{appengine})
to avoid lost application state and/or shorten pause times.

\tool triggers conversions automatically as data is accessed. 
To track its progress, \tool attaches a version
identifier to the value of each entry, converting only those
keys/values that are out of date. Conversions are written by the
developer. \tool ensures updates are installed atomically in a way
that supports fault tolerance. \tool also automatically ensures that conversions
take place atomically with the triggering database action; as such,
\tool avoids races that could clobber concurrent accesses. \tool
requires a conversion function to only access the corresponding old
value/key, not \emph{several} old key/values; to allow otherwise
could violate logical consistency depending on the order that
conversions are triggered. To support laziness, transformations to
keys must be reversible and unambiguous.
Examination of open-source software histories, and our own experience,
suggests that realistic conversions typically satisfy these
restrictions. 

We describe a proof-of-concept implementation of \tool as an extension to the popular \redis
key-value store.
We evaluate this implementation extensively, using both micro-benchmarks (the
standard \redis performance benchmark) and macro-benchmarks (two feature-rich applications, \redisFS and Amico). 
%
Our experiments suggest that \tool imposes essentially no overhead during normal operation and that complex applications can be upgraded with zero downtime.
In particular, when upgrading \redisFS we used \tool to upgrade the
filesystem data, and Kitsune~\cite{DBLP:journals/toplas/HaydenSSHF14}, a whole-program
updating framework for C, to dynamically update the \redisFS
driver. As a result, we could seamlessly maintain the file system
mount point during the upgrade, resulting in zero downtime.


In summary, we make three contributions: 

\begin{itemize}
\item We identify the challenges for evolving NoSQL databases without
  downtime (Section~\ref{sec:overview}) and, to our knowledge, we
  propose the first general-purpose, automatic solution to this problem (Section~\ref{sec:upgrades}). 

\item We describe a proof-of-concept implementation as an extension of the Redis key-value store (Section~\ref{sec:impl}).

\item We evaluate this implementation extensively, and we show how to combine \tool with a dynamic program updating tool for zero-downtime upgrades (Section~\ref{sec:experiments}).
\end{itemize}


\section{The problem with on-line upgrades}
\label{sec:overview}

This section details the problem of updating a NoSQL database on-line,
and the drawbacks of prior solutions. Our approach, KVolve, is
detailed in the next two sections.

\subsection{NoSQL DBs and KV stores}\label{sec:background} 

NoSQL databases distinguish themselves from traditional \emph{relational
  database management systems} (RDBMSs), by supporting a simple,
lightweight interface. Our focus is on a NoSQL variant referred to as
a \emph{key-value (KV) store} which, as the name implies, focuses on
mapping keys to values. 
There are two core operations: \code{GET} $k$, which
returns the value $v$ to which $k$ maps in the database (or ``none''
if none is present); and \code{SET} $k\; v$, which adds (or
overwrites) the mapping $k \rightarrow v$ in the database. 
Example KV stores include \redis (the most
popular~\cite{redis}, and the target of our proof-of-concept implementation), Project
Voldemort~\cite{vol}, Berkeley DB~\cite{bdb}, and many
others~\cite{others}. 

While a KV
store may place no formatting requirements on values (i.e., treating
them as bytearrays), applications typically store values adhering to
formats such as JSON~\cite{json}, Avro~\cite{avro}, or Protobufs~\cite{protobuf}. Some KV stores do expect a
specific value format; e.g., Cassandra defines typed ``rows'' in
``tables,'' and MongoDB employs ``documents.'' Likewise, key formats
may be unstructured (i.e., just strings) or have some structure added
by the system (e.g., a notion of \emph{prefix}, or \emph{namespace}).



\subsection{Example application and update}
\label{sec:basicexample}

As an example (adapted from Sadalage and Fowler~\cite{sadalage}),
consider an on-line store which keeps track of purchase
orders. The application stores these orders in a KV store, using keys
of the form \code{order:}$n$, where $n$ is a unique invoice number,
and values formatted as JSON records describing the purchasing
customer and what was ordered. In this key, \code{order} is a prefix
to assist in key grouping, e.g., as part of the encoding of a
table. An example JSON record is shown in
Figure~\ref{fig:purchase}(a).\footnote{JSON defines four
  primitive types: numbers, strings, booleans, and null. It also
  defines two container types: arrays, which are an ordered list of
  values of the same JSON type; and objects, which are an unordered
  collection of values of any JSON type, with field labels. We use
  JSON as an example only; other formats are also supported by
  KVolve.}


Suppose we wish to upgrade the application to support
differentiated pricing, which necessitates changing the data format in
the KV store. Keys remain the same, but values change: we rename
the field \code{price} to \code{fullprice}, and insert a new field
named \code{discountedPrice} that is a possible reduction of the original
price. The updated \code{orderItems} array (the last element of the
JSON object) is shown in Figure~\ref{fig:purchase}(b).

\begin{figure}
\begin{minipage}{.9\columnwidth}
\begin{lstlisting}[showstringspaces=false,numbers=none]
{ "_id": "4BD8AE97C47016442AF4A580",
  "customerid": 99999,
  "name": "Foo Sushi Inc",
  "since": "12/12/2012",
  "order": {
    "orderid": "UXWE-122012",
    "orderdate": "12/12/2001",
\end{lstlisting}
\vspace*{-1em}
\end{minipage} 
\begin{minipage}[t]{.45\columnwidth}
\begin{lstlisting}[showstringspaces=false,numbers=none]
    @"orderItems": [
       { "product": "Cookies",
         "price": 19.99 }
    ]@
\end{lstlisting}
\vspace*{-1em}
\end{minipage}
\vrule \hfill
\begin{minipage}[t]{.5\columnwidth}
\begin{lstlisting}[showstringspaces=false,numbers=none]
@"orderItems": [
  { "product": "Cookies",
    "fullPrice": 19.99,
    "discountedPrice": 16.99 }
]@
\end{lstlisting}
\vspace*{-1em}
\end{minipage}
\begin{lstlisting}[showstringspaces=false,numbers=none]
} }
\end{lstlisting}
\centering
\begin{tabular}{p{.45\columnwidth}p{.45\columnwidth}}
(a) original format & (b) updated format \\
\end{tabular}
\caption{JSON object, and update}
\label{fig:purchase}
\vspace{-5 mm}
\end{figure}

\subsection{Past approaches to on-line data upgrades}

\mypara{Eager, stop-the-world data upgrades}
One approach for implementing the data upgrade described above is
to simply halt all client applications and use a script to convert all
the data in the KV store that is out of date. Once all the data is
updated, the clients can be restarted. For our example, the conversion
script would get each purchase order value, modify it, and store back
the updated result. In particular, for each existing purchase order
value, the script would replace the existing elements of the
\code{orderItems} array with new elements whose \code{price} field
is renamed \code{fullPrice}, and which contain with a new
\code{discountedPrice} field initialized to the old \code{price}
value.

While simple, the downside of this deployment strategy is the disruption in service
while the database is being upgraded. Our experiments show that for even
modest-sized databases (hundreds of thousands of keys), this disruption
can be on the order of several minutes. 
On a larger scale, Twitter
has deployed 6,000 instances of \redis at each of many data centers, 
with instances using 40TB heaps~\cite{twitter}. 
Any update to that data of that size will be very time consuming.
%

\mypara{Manual, lazy data upgrades}
Rather than pause service while the database is being \emph{eagerly}
upgraded, Sadalage and Fowler suggest that the programmer can modify
the new version's code to handle old and new formats, and migrate old
data \emph{lazily}, when it is encountered. For this example, the
application could try to access the \code{fullPrice} field of a
purchase order's \code{orderItems} array. If that field is not
present, the application can update the value as described above, and
then try again.

This approach works but adds a greater burden on the programmer, who
must add the version checking code, and code to upgrade out-of-date
values. Such upgrades imply that what once was a \emph{\code{GET}
  may now involve an additional \code{SET}} to the updated format.  As
we explain in the next subsection, this added operation could result
in data-corrupting race conditions. Worse, version checking/updating
code will grow over time, as mentioned in Section~\ref{sec:intro}, and
will become more complex as applications expand to deal with a variety
of data values. The end result is a significant maintenance headache.

\section{\tool}\label{sec:upgrades}
\label{sec:basics}

\begin{figure}
\centering
\scalebox{.6}{\includegraphics{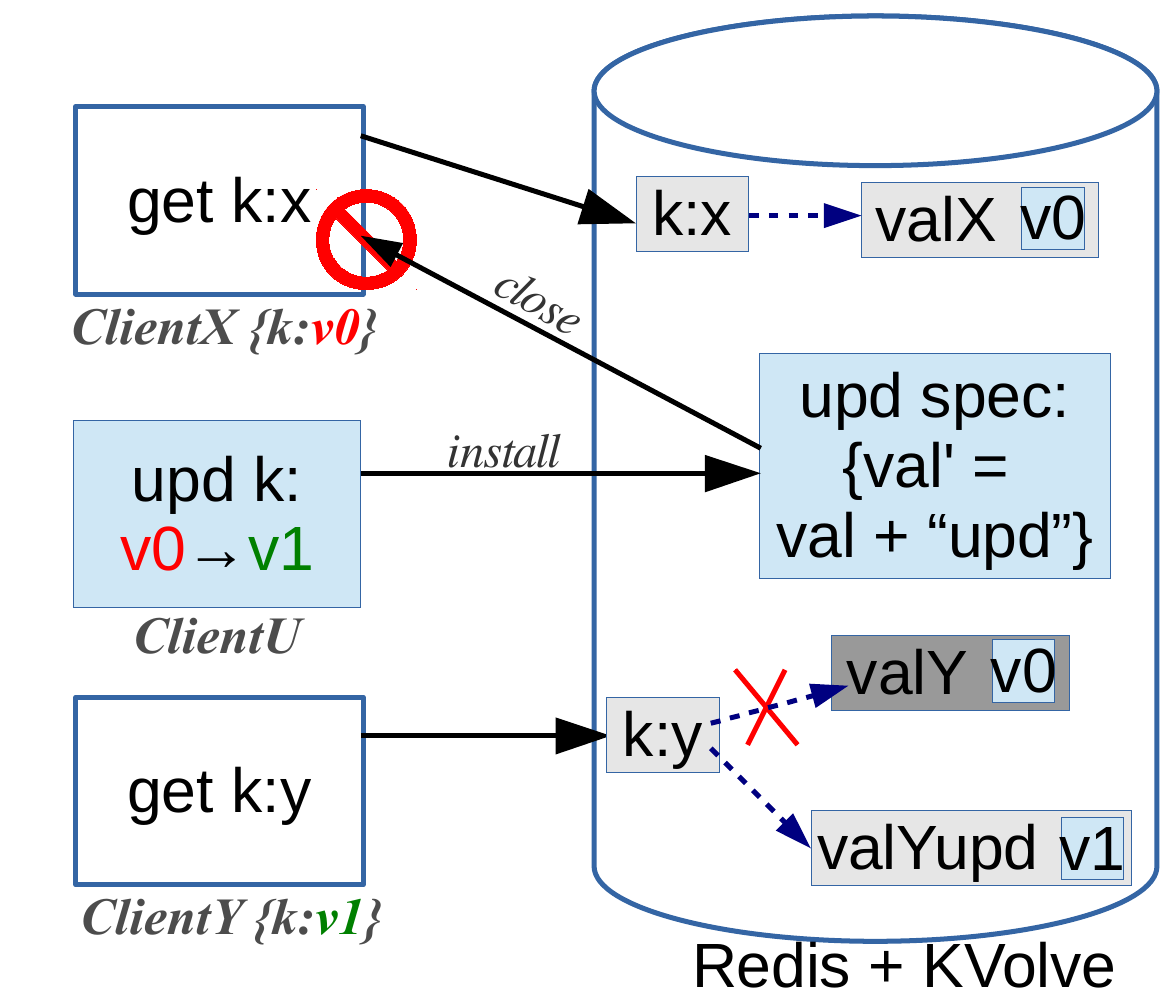}}
\caption{\tool architecture.}
\vspace{-5 mm}
\label{fig:arch}
\end{figure}

\tool aims to solve the on-line upgrade problem in a way that enjoys the
best features of lazy and eager data upgrades.
In particular, \tool migrates data lazily, as it is accessed
by applications, thus eliminating any long, disruptive pause. But \tool
presents a \emph{logically consistent view} to applications, 
providing the \emph{appearance} that all the data is instantly upgraded to the new version.
As a result, programmers do not need to add any
version-management code to their applications; they simply write the
application assuming the most recent data version. 
Because the lazy
migration is handled by \tool, it can ensure there are no errors due
to concurrent interactions.

\subsection{KVolve design}

Our approach is characterized by three techniques. 

\mypara{Versioned data} 
We associate logical version identifiers with the database
content. Rather than having a global data version, we track separate
versions for data associated with different key prefixes. E.g., data
mapped from keys $p_1$:$x$ (for all $x$) has a separate version ID
space than data mapped from keys $p_2$:$x$ (for all $x$). A version
tag is stored with each data item indicating its actual version, which
might be earlier than its logical one (i.e., if the data item has not
been migrated yet). Version tags are invisible to applications
accessing the database. When the application connects to the database,
it indicates
the version IDs of the key prefixes it will use, and \tool compares
them to the logical versions of those prefixes.  If the two IDs match,
\tool accepts the connection.


\mypara{Update specifications and state transformer functions}
When the database is to be upgraded, the operator installs a
specification describing the mechanics. In particular, the specification
defines the new logical prefix versions, and provides \emph{state transformer}
functions to be used to upgrade particular values. Each transformer
function $f$ is associated with a key prefix $p$. If a key of the form 
$p$:$x$ (for some $x$) maps to a value $v$, then $p$:$x$ will be
updated to map to $f(v)$. KVolve can handle key format changes, too,
as discussed below.

\mypara{On-demand (lazy) transformation}
Once the update specification is installed, applications connected to
the database that are out of date must be disconnected. They will
reconnect at the new code version (mirroring the situation with eager
upgrades). Doing this is not onerous for most applications as discussed
in Section~\ref{sec:manual}.

Once a new application version starts running, it will
submit \code{GET}s and \code{SET}s to KVolve for handling. If a \code{GET} accesses a
value that is out of date, KVolve first updates the stale item using
the appropriate transformer function. If a data item is several
versions out-of-date, transformer functions will be composed
and applied automatically. 

We illustrate these three techniques in Figure~\ref{fig:arch}. Here,
\emph{ClientX} initially connects at version \emph{v0}, and is able to
access the value mapped to from $k$:$x$ safely, since it is also at
\emph{v0}. Then, \emph{ClientU} updates the database to version
\emph{v1}, and includes a state transformer function for prefix
$k$. This function concatenates the string ``upd'' to an existing
key's value. This update causes \emph{ClientX} to be disconnected
because its version \emph{v0} is now inconsistent with the database's
logical version \emph{v1}. Finally, \emph{ClientY} connects to the
database at version \emph{v1}. It performs a \code{GET} on key $k$:$y$. This
key maps to a stale value, having version \emph{v0}. Therefore, KVolve
remaps the key to a the value produced by running the 
transformer function on the old value. Then it returns the updated
value to \emph{ClientY}.


\subsection{Ensuring logical consistency}
\label{sec:logical}

KVolve's goal is to provide a logically consistent view to
applications. That is, any sequence of commands issued by up-to-date
clients should produce the same results whether interacting with a
fully (i.e., eagerly) updated database or with one whose data is being
migrated lazily, as it is accessed. This goal imposes three
requirements on KVolve's implementation. First, state transformations
must occur atomically with the operation that induced them. Second,
transformer functions may only reference the old version of the
to-be-updated key/value, and key changes are restricted. Third, the
update specification must persist once it is installed, so
that logical consistency is maintained following recovery from a
fault.

\mypara{Atomicity} Upgrading data atomically ensures that
clients accessing the data concurrently with the lazy transformations 
will not 
cause anomalies that cannot occur with eager upgrades or during normal operation. 
To see
how an anomaly could be introduced, consider a trace with a \code{GET}
$k$:$x$ by client $A$ and a \code{SET} $k$:$x$ $w$ by client
$B$.
Suppose that $k$:$x$
maps to $v$ in the old database, and the update's transformer function
$f$ operates on a key's old value to produce the new one.

In an eager update, $k$:$x$'s value $v$ is updated to be $f(v)$. Then
there are two possible execution schedules: client $A$ could retrieve $f(v)$ and
client $B$ could set $k$:$x$ to $w$, or client $B$ could do the set,
in which case $A$ returns $w$ (which is already up to date). 
In both cases, the final database maps $k$:$x$ to $w$.

In a lazy update, a transformer must be invoked before returning the value
to $A$. One way to implement this would be to convert client $A$'s \code{GET}
into two commands when dealing with an out-of-date value: \code{SET} $k$:$x$
$f(v)$ (i.e., set it to the updated value) and then \code{GET} $k$:$x$ (i.e.,
return that transformed value $f(v)$ to the client). But in this case,
client $B$'s \code{SET} could be scheduled in between client $A$'s \code{SET} and
\code{GET}. This would result in $A$'s \code{GET} returning the $w$
\code{SET} by $B$, but then $A$'s \code{SET}
overwriting $w$ with $f(v)$. This final state would never be possible
in the eager case. 
Effectively, improperly implemented lazy updates could cause client $B$'s operations to
fail silently, without notifying $B$ of the failure. 
This anomaly violates logical consistency. In contrast, this
scheduling is not possible if client $A$'s read and update must always
be atomic. One of KVolve's benefits, over by-hand modification of code
to support lazy migration, is that it can ensure atomicity
automatically. 

\mypara{Limited domain of transformer function} KVolve restricts
transformer functions $f$ in two ways. First, the transformer may only
operate over the old version of the key/value it is updating, and not
any other items in the database. Second, the transformer may not
change keys arbitrarily; instead it only supports unambiguous
bijections on a key's prefix.

The first restriction ensures that when $f$ runs, it will operate on
the same keys/values it would have if run when the update was
installed. This is because only the first \code{GET} of a key could
possibly see a stale
value, and it will immediately update it. As such, it ensures a
logically consistent view. On the other hand, if we allowed $f$ to
access other data items, it is easy to see how logical consistency is
broken. For example, suppose the function $f$ to update a key
$k$:$x$'s value also examined $m$:$x$'s value $v$. If the new-version
code executed a \code{SET} $m$:$x$ $w$ prior to a \code{GET} $k$:$x$,
then $f$ would read $w$, not $v$. 

The second restriction ensures that lazy key updates can be
implemented safely and efficiently. After an update is initiated, the
new application version will issue commands using the \emph{new}
keys. For example, suppose an update changes the prefix from \code{k}
to \code{m:j}, so that keys \code{k:}$n$ would become \code{m:j:}$n$
(for all $n$). After the update, applications will submit commands
like \code{GET m:j:}$n$. If the key is present, we need to be sure
that it is a new-version key, not an old one that has yet to be
transformed; as such transformations may not map to key prefixes that
are also present in the old database version. On the other hand, if
the key \code{m:j:}$n$ is not present, KVolve should look for the
\emph{old} version of the key, in case it is there and thus needs to
be updated. To do this, KVolve will have to run the transformation
backwards, i.e., on \code{m:j:}$n$ to produce \code{k:}$n$.  
Limiting transformations to key prefixes helps make backward transformation
efficient, since KVolve can match keys against (new-version) prefixes
directly. 

Restricting transformation functions in this way is conceptually
limiting, but not practically so, we believe. We analyzed
18 
of the most active projects on GitHub that used \redis to
store program data, and none of the programs contained value changes
that were dependent on other value changes, and key changes were
limited to prefix changes.

\mypara{Fault tolerance}
Many KV stores provide fault tolerance guarantees; i.e., there is a way to
checkpoint the database so that it can be recovered after a crash. As
such, if the database crashes during a lazy upgrade, KVolve's
implementation should ensure the logical view is retained following
recovery. KVolve ensures this by (a) storing per-data version tags in
the database, so they are made persistent; and (b) storing the update
specification (and logical version) in the database, atomically, when
the update is installed. This way, if the database crashes before the
update is fully installed, then on recovery the database will still
appear (correctly) to be at the old version. But once the update is
fully installed, then the database identifies as being at the next
logical version, and lazy migration can pick up where it left off
after recovering from a failure.

\section{Implementation}\label{sec:impl}


This section describes our implementation of \tool as a modular
extension to the popular \redis key-value store.

\subsection{\tool implementation overview}\label{sec:changes}

\begin{figure}[t]
  \begin{centering}
  \includegraphics[scale=0.65]{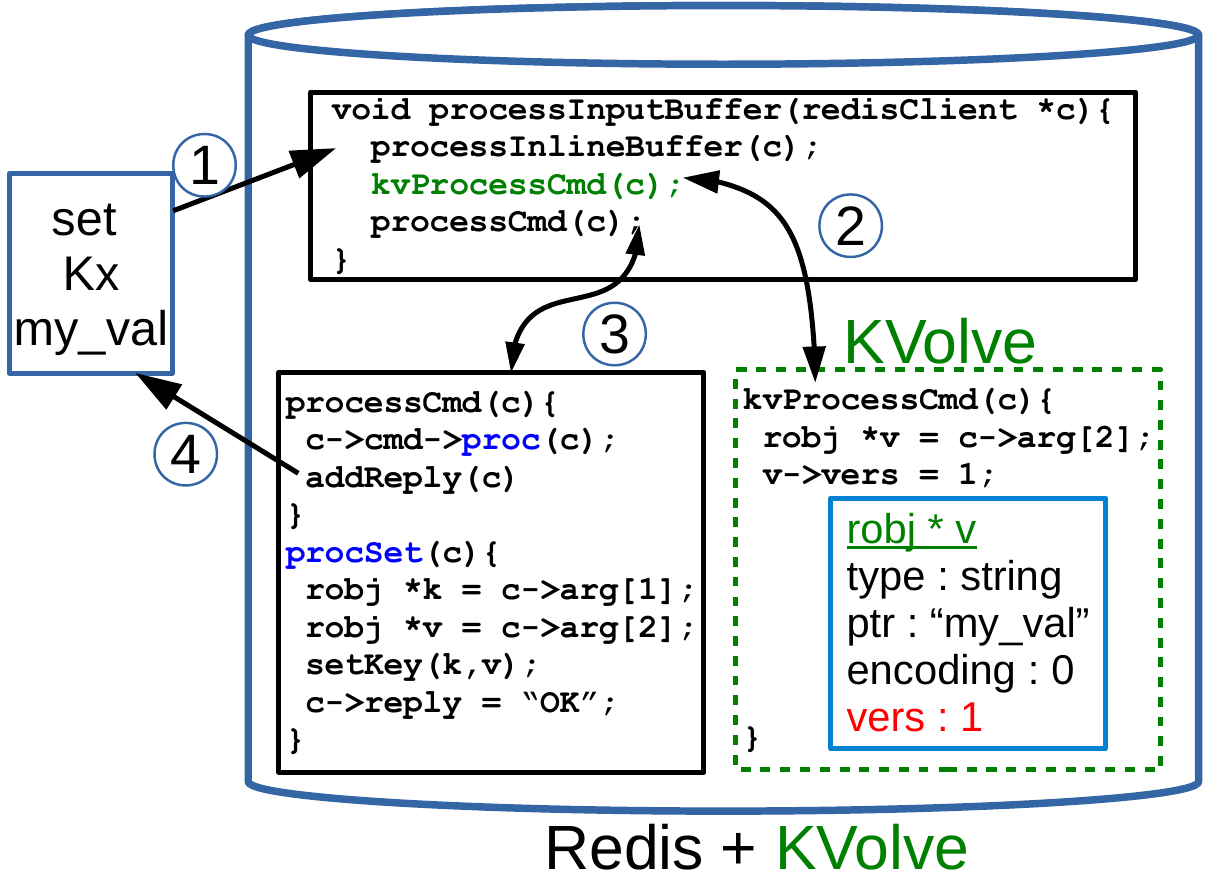}
  \caption{Workflow for \redis and \tool}
  \label{fig:arch2}
  \end{centering}
\vspace{-5 mm}
\end{figure}

\tool is implemented as a separate library compiled into \redis.
It works by preprocessing commands coming in from the client before passing
them along to \redis, as depicted in Figure~\ref{fig:arch2}. In Step
1, the client issues the command, e.g., \code{SET Kx} \code{my_val}.  
In Step 2, \code{kvProcessCmd}, \tool's hook, is called to preprocess
the command (the dashed green box is the \tool library).  Once the
\tool preprocessing is complete (which might involve changes to
data's contents and version field), control returns to normal
\redis. In Step 3, \redis's \code{processCmd} function calls the
function pointer shown in blue (which depends on the choice of
command---here it is \code{procSet} because the client requested
\code{SET}), and this adds the affected object to the database,
including any changes to the version field set during \tool's
processing.  Finally, in Step 4 \redis responds to the client's
request, acknowledging to the client that it successfully executed the
\code{SET} command.

All of this is sure to be atomic because \redis is single-threaded: it
processes each command it receives in its entirety before moving to
the next. \redis provides commands, such as \code{multi}, that can be
used to execute a group of commands atomically; \tool's design works
in concert with such commands. We also believe that \tool's basic
``interceptor'' architecture would work in multi-threaded KV store
implementations by employing appropriate synchronization.

\subsection{Describing and installing updates}\label{sec:manual}

An update consists of \emph{transformer functions} that will convert
the old version of a key and/or value to the new version. The
programmer compiles the transformer functions into a shared object
file that she can direct \tool to install (using a repurposed \redis
command). Once installation is complete, the shared object (and
metadata about it) is stored persistently in \redis.

There are two kinds of updates: key/value updates and key
updates (only). As an example of the former, Figure~\ref{fig:jsonmu}
shows a transformer for the example from Figure~\ref{fig:purchase}.
The old key (a string) and value (binary data) are passed in by
reference, and the function will update them to the new versions via
these references. In this case, the body of the function uses the
Jansson library~\cite{jansson} to implement the change to the purchase
order example from Figure~\ref{fig:purchase} described in
Section~\ref{sec:basicexample}; the last two lines update the value
(the key is not changed). Writing this code is a bit tedious. As done
in our prior work~\cite{DBLP:journals/toplas/HaydenSSHF14,
  saur15morpheus,DBLP:journals/corr/SaurDH15}, we could easily implement a 
domain-specific language to simplify the process. 


\begin{figure}
\begin{lstlisting}
void test_fun_updval(char ** key, void ** value, size_t * val_len){
  json_t *root = json_loads(*value);
  json_t *arr = json_get(json_get(root, "order"), "orderItems");
  // iterate over the {product: ... , price: ... } array entries
  for(i = 0; i < json_array_size(arr); i++){
     // Set discountedPrice to (price - 3)
     // Set fullPrice to price, then delete price
  }
  *value = json_dumps(root); // Set the updated value/**\label{line:valset}*/
  *val_len = strlen(*value); // Set the updated val length/**\label{line:lenset}*/
} 
\end{lstlisting}
\caption{Example pseudocode transformer function for JSON}
\label{fig:jsonmu}
\vspace{-5 mm}
\end{figure}

Along with the transformer functions, an update specification contains
a function that is invoked when the shared object is loaded as part of
an update. This function consists of a series of calls to install
transformer functions. Our example above is installed by the following
call:
\vspace{-.2mm}
\begin{lstlisting}[showstringspaces=false,numbers=none,language=C]
kvolve_upd_spec("order","order", 0, 1, 1,test_fun_updval);
\end{lstlisting}
\vspace{-.2mm} 
This call indicates that the \code{order} prefix doesn't change, from
version 0 to version 1, while the \code{test_fun_updval} should be
called for each key with the prefix \code{order}.

Key prefixes can be changed without requiring a transformer function.
For example, in the Amico program described in
Section~\ref{sec:amico}, the keys are renamed from the prefix
\code{amico:followers} to the prefix \code{amico:followers:default}.
To describe this update, the initialization function would include the call:
\vspace{-.2mm}
\begin{lstlisting}[showstringspaces=false,numbers=none]
kvolve_upd_spec("amico:followers", "amico:followers:default",1, 2, 0);
\end{lstlisting}
\vspace{-.2mm}
where the version numbers are 1 and 2, and the final 0 indicates that
there are no functions to manipulate the value.


\tool will close the connection to all
clients using the old version of updated prefix(es). A disconnected
client will not be allowed to reconnect until it is upgraded to the
new version. Clients not using updated prefix(es) will not be affected.
To use \tool, therefore, processes connecting to \tool must be
coded to support disconnection, upgrade, and restart. This is straightforward, in our experience,
and need not be disruptive to end users. For example, customer-facing
clients in web browsers can maintain session-permanence even as the
backend servers, i.e., those connected to a \tool DB, are upgraded. Such update patterns
are common with load-balancing stateless servers~\cite{oracle}.


\tool stores update specifications indefinitely. We find that the
transformer functions take up a small amount of space relative to the
rest of the data. 
However, if program updates are very large or very frequent, one could
employ a background client or similar thread to force updates to outdated data by
\code{GET}ting them all; once done, all update entries could be
freed. (This would essentially be a hybrid of the lazy and eager approach.)

\subsection{Key lookup}\label{sec:keylookup}


After an update is installed, the database's logical version is
advanced. Because the new version might transform the format of keys,
KVolve may need to look up the \emph{old} version of a key specified
in an application's \code{GET} or \code{SET} commands, so that it can
update that key (as per Section~\ref{sec:logical}). To support this,
we use a \emph{update information hash table} (UIHT). 
This table maps a key prefix to a record which contains both that same
prefix and pointers to records that describe the next and/or previous
versions of the prefix. For example, after a key update from
\code{foo:} to \code{foo:bar:}, the table would map prefix
\code{foo:} to a record $q$ whose next pointer would be to a record
$r$ for \code{foo:bar:}, which points back to $q$; the table
would also map prefix \code{foo:bar:} to $r$. As such \tool can trace
through all current and former versions of a 
prefix for applying updates. The table's records also contain
transformation functions for moving forward between versions, and
track the IDs of client connections that are using a particular prefix
version, so they can be disconnected on an update to it.

After a key update, client queries will use the new key format; e.g.,
after updating prefix \code{foo:} \code{foo:bar:}, client commands
will refer to keys \code{foo:bar:}$n$. \tool will first look to see if
a key exists under the issued name. If it is not there, an old,
non-updated version of the key may be present.  As such, \tool looks
up \code{foo:bar:} in the UIHT to see if a record is present that maps
to an old prefix. In this case, \tool will see that prefix
\code{foo:bar:} points back to prefix \code{foo:}, so it reissues the
client command with key \code{foo:}$n$ instead. If it finds it, it
updates the key name to \code{foo:bar:}$n$ and returns the value to
the client. If no match is found, it will continue to follow
backpointers in the UIHT if prior versions should be considered. If
none are found, meaning that no key is present, \tool returns control
to \redis without further action.

Looking for keys under previous prefixes adds
additional lookups only once (during the update) when the key is present under an
old version.  However, the case where there is no key present under any prefix
version will add unnecessary additional lookups each time the non-present key
is queried. An application that frequently queries keys that are not present
where there has been a prefix change could negatively impact performance. In
previous work~\cite{DBLP:journals/corr/SaurDH15}, we experimented with adding
a \emph{sentinel value} to mark the key as absent, skipping the step of
checking for the key under previous prefixes, thus saving on lookup time but
adding a bit extra storage.

\subsection{Getting and setting values}

\begin{figure}[t]
  \begin{centering}
  \includegraphics[scale=0.65]{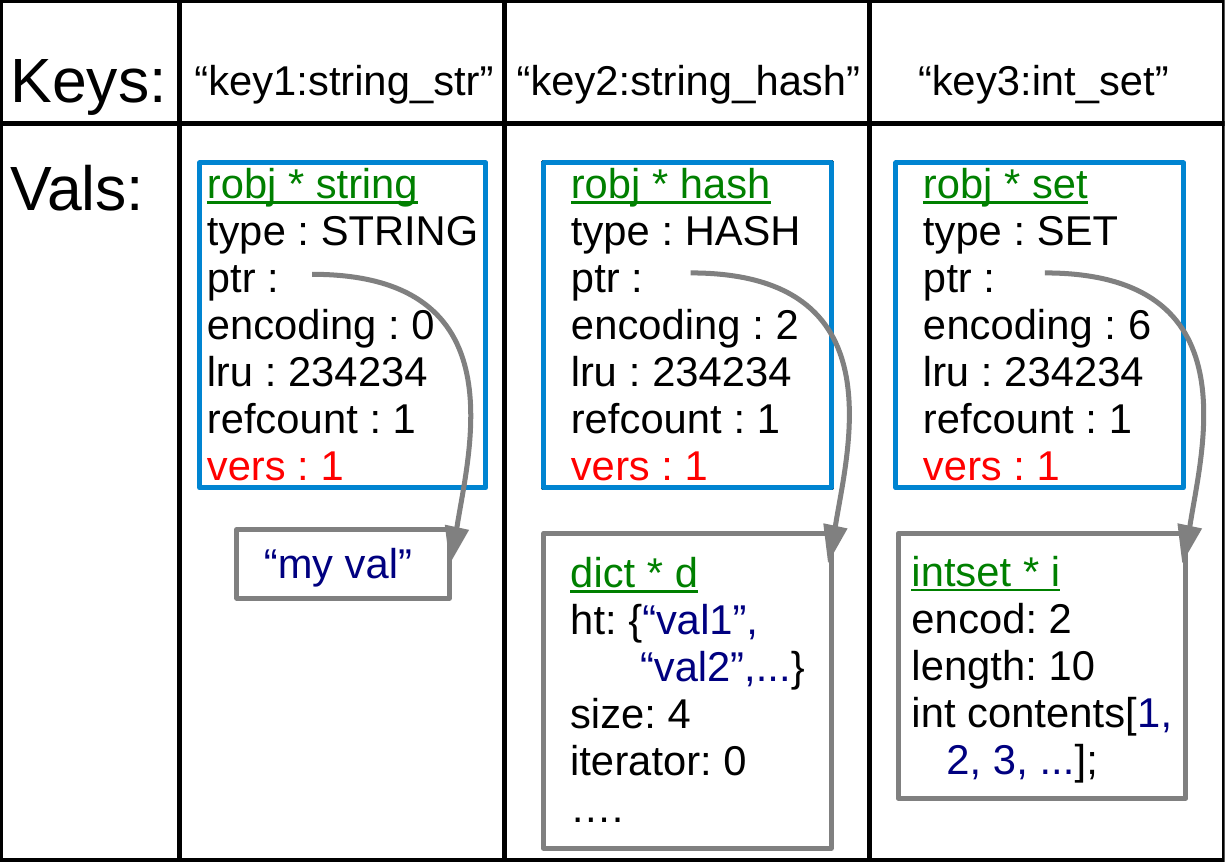}
  \caption{Storing different data types}
  \label{fig:robjs}
  \end{centering}
\vspace{-5 mm}
\end{figure}


In the simplest case, keys map to string values. We consider this case
first.
Our implementation currently supports 36 \redis
commands and all of the main \redis data
structures (string, set, list, hash, sorted set); we discuss
containers in the next subsection.
%
We focused on implementing commands that modify data. The majority of
\redis' commands that we did not implement do not impact updating the data
(e.g. commands related to networking such as the pub/sub functionality or connectivity).
%

%

\mypara{GET}\label{sec:get_triggers_update}
If the client request involves getting a string,
\tool must first prelookup the existing \code{robj} value structure in
the database to get the version information.  An example of the key-value
pair for string types is shown for \code{key1:string_str} in the first
column of Figure~\ref{fig:robjs}.
This action retrieves a pointer to the actual
object structure that is stored in the database, so any modifications that \tool makes
to this object will be automatically stored in \redis. 
Note that this requires an additional database lookup by \tool, on top of
the one that \redis will do later when it does its own lookup to handle the client request.
However, this is an O(1) operation and does not incur excessive overhead
relative to the other operations that \tool must already perform.

If the version field of the \code{robj} (in this example the version is
\code{1} shown in red for \code{key1:string_str} in Figure~\ref{fig:robjs}) is
current for the prefix of the key, or if the key is not present under the
current or any former prefix (and therefore no \code{robj} exists for the
key), then \tool returns control to \redis and does no further processing.  If
the version is not current, either in the current prefix or a former
prefix, \tool will update the key and value, as specified. All of the
necessary information to perform the update (the transformer functions themselves, and
the meta-data about which prefixes and versions the updates apply to) is
stored in the update information hash table, and \tool uses that information to apply the 
update as follows:

\begin{itemize}

\item In the case of a key prefix change, \tool uses the update information
from the hash table to perform the key rename, leaving the value
untouched. 
%

\item In the case of a value change, \tool calls any applicable user-supplied
functions and applies them to the value, starting from the oldest needed update
and working forward to the current version. After all of the transformer functions have been applied, \tool
stores the updated value in the \code{robj} (which is a pointer to the actual
structure stored in the database) and updates the version string to match the
current version by setting the field in the \code{robj}. 
\end{itemize}
If both actions (key prefix and value change) are necessary, \tool will perform both.
\tool then returns
control flow to \redis, and when \redis performs its own \code{GET}, it will
retrieve and return the newly updated key to the client.

\mypara{SET}  If the client request involves setting a string,
\tool first checks to see if the request has any flags that
would prevent the value from getting set.  These flags, \code{XX} or \code{NX},
respectively specify to only set the key if it already exists, or only set the
key if it does not already exist.  If necessary, \tool does a lookup in the
database to determine if the key exists, indicating if the value will be set
for the requested key. (As described in Section~\ref{sec:changes}, if the
prefix changes, \tool will search for the key under the old prefix to see if it exists.)
 If the value will not be set due to the flags, \tool
does nothing and returns control to \redis.  In this \code{SET} command, or
any such command where \redis will be adding the \code{robj} to the
database, \redis deletes the old \code{robj} and replaces it with the new one
from the client's request.  Therefore, all that \tool must do is set the
most current version string in the \code{robj} for the prefix of the key.
(Remember that there is no need to attempt to update the value in the key, because
the client's provided value is guaranteed to be at the up-to-date version.)  

If this set occurs after a key prefix change, \tool must delete the old value
for the key to ensure that deprecated key versions are not unnecessarily
bloating \redis.  For example, in a change to \redisFS (presented in
Section~\ref{sec:redisfs}), an old key prefix was named \code{skx:/} but
after an update, the new name postfixes \code{DIR} such that the key is
now named \code{skx:DIR:/}.  If the user were to set the key \code{skx:DIR:/root}
before getting (and updating) it, this would leave the old key \code{skx:/root}
still in the database.  Therefore, \tool must check to see if the existing
version under the old key prefix exists, and if it does, delete it.  It does
this by first checking if the prefix had any previous changes.  If
not, it does nothing.  If so, it checks and deletes the old key if necessary.
At this point, \tool returns control to \redis, and \redis adds the \code{robj}
structure to the database, which  also contains the
updated version string to be retrieved later if necessary. 

\subsection{Sets, hashes, lists, and sorted sets}\label{sec:containers}
The other \redis value data structures are containers of sub-values.
The base of \redis containers are all \code{robj} structures, and they
store the actual data.  
Figure~\ref{fig:robjs} shows examples
in columns two and three of \code{robj}s that contain a hash of strings and a
set of integers, respectively. 
\tool stores version information in the container, not in the
contained values (to avoid more pervasive changes to \redis), so
updates to containers happen all at once.

The process for doing a \code{GET} or \code{SET} on one of the container elements is 
the same as for the string type described in
Section~\ref{sec:get_triggers_update}, except that if an update is
necessary then all sub-elements are updated, not just one, 
using a \redis-provided iterator.  

\section{Experimental results} \label{sec:experiments}

This section considers the performance impact of \tool, during normal
operation and during an update. Our experimental results are summarized
as follows:
\begin{itemize}
\item Using the standard benchmark that is included with \redis, we found that
   \tool adds essentially no overhead during normal operation, and we determined
   that storing the version and update information in \redis adds only about a 
   15\% overhead in space.

\item We updated the \redisFS file system which included renaming some keys and
   compressing some data stored in keys, and found the operating overhead to be in noise,
   and the pause time to be close to zero as opposed to 12 seconds for
   an offline data migration.

\item We updated the Amico social network system and found no added overhead,
   with a pause time of close to zero as opposed to 87 seconds for
   an offline data migration.

\end{itemize}
All experiments were performed on a computer
with 24 processors (Intel(R) Xeon(R) CPU E5-2430 0 @ 2.20GHz) and 32 GB RAM
with GCC 4.4.7 on Red Hat Enterprise Linux Server release 6.5.  All tests
report the median of 11 trials, and communication was via localhost with 
\tid.03 ms latency.


\subsection{Steady state overhead}\label{sec:exp1}

\begin{table*}[t!]
\begin{center}
\centering
\label{my-label}
\caption{\bench  for \redis vs \tool (times in seconds, median of 11 trials)}
\begin{tabular}{|l|l|ll|lll|lll|lll|}
\hline
&           & \multicolumn{2}{c|}{\redis} & \multicolumn{3}{c|}{No NS, \tool} & \multicolumn{3}{c|}{With NS, \tool} & \multicolumn{3}{c|}{\&Prev NS, \tool}  \\ \cline{3-13}
&           & time   & siqr        & time  & siqr        & OH               & time   & siqr        & OH               & time  & siqr        & OH               \\ \thickhline
\textbf{With single}  &String Get & 58.83  & \siqr{0.25} & 58.08 & \siqr{0.11} & \textbf{-0.77\%} & 58.46  & \siqr{0.26} & \textbf{-0.12\%} & 59.51 & \siqr{0.85} & \textbf{1.67\%}  \\ 
\textbf{instructions} &String Set & 63.52  & \siqr{1.04} & 63.66 & \siqr{0.70} & \textbf{0.22\%}  & 65.39  & \siqr{1.49} & \textbf{2.49\%}  & 64.82 & \siqr{0.35} & \textbf{2.05\%}  \\ 
             &List Pop   & 58.47  & \siqr{0.41} & 58.93 & \siqr{0.68} & \textbf{0.79\%}  & 58.71  & \siqr{0.15} & \textbf{0.41\%}  & 59.34 & \siqr{0.47} & \textbf{1.49\%}  \\ 
             & List Push  & 59.49  & \siqr{0.66} & 59.55 & \siqr{0.56} & \textbf{0.10\%}  & 60.02  & \siqr{0.74} & \textbf{0.89\%}  & 60.87 & \siqr{0.94} & \textbf{2.32\%}  \\ 

\thickhline

\textbf{With 10}        &String Get & 9.73  & \siqr{0.30} & 9.75  & \siqr{0.27} & \textbf{0.21\%}  & 9.96   & \siqr{0.25} & \textbf{2.36\%} & 9.93  & \siqr{0.20} & \textbf{2.06\%}  \\ 
\textbf{\textit{pipelined}}      &String Set & 13.77 & \siqr{0.26} & 14.56 & \siqr{0.18} & \textbf{5.74\%}  & 14.56  & \siqr{0.32} & \textbf{5.74\%} & 14.48 & \siqr{0.33} & \textbf{5.16\%}  \\ 
\textbf{instructions}   &List Pop   & 9.60  & \siqr{0.32} & 9.71  & \siqr{0.25} & \textbf{1.15\%}  & 9.55   & \siqr{0.43} & \textbf{-0.52\%}& 9.63  & \siqr{0.27} & \textbf{0.31\%}  \\ 
               &List Push  & 14.22 & \siqr{0.39} & 14.38 & \siqr{0.25} & \textbf{1.13\%}  & 14.40  & \siqr{0.41} & \textbf{1.27\%} & 14.48 & \siqr{0.36} & \textbf{1.83\%}  \\

\hline 
\end{tabular}

\vspace{-4 mm}
\label{tab:overhead}
\end{center}
\end{table*}

\begin{table}[]
\centering
\caption{Max resident set size (RSS)}
\begin{tabular}{|l|r|}
\hline
\textbf{Program}      & \textbf{Max RSS} \\ \hline
\redis, empty             & 7.7MB       \\ \hline
\redis, 1M 10-byte values    & 112.1MB      \\ \hline
\tool, empty             & 7.7MB       \\ \hline
\tool, 5 prefixes, 1M 10-byte values & 128.6MB     \\ \hline
\end{tabular}
\label{tab:rss}
\vspace{-4 mm}
\end{table}

First we report the steady state overhead for \tool reported by \redis's
included benchmark, \emph{\bench}. \bench acts as a client that repeatedly
issues commands to \redis. The default settings for \bench are with 50 clients,
with 10,000 repetitions of a single operation at a time (only 1 request per round
trip), and with a single key (getting or setting a single key multiple times).
However, \bench allows many different configurations. For a longer benchmark, we increased 
the number of operations to 5 million operations and for a more realistic benchmark
we performed these operations over 1 million keys, 
leaving the rest of the default settings alone. We ran this
experiment over localhost which had a latency of \tid.03ms.  We chose three types
of \code{GET} operations (string gets, set pops, and list pops) and three types
of \code{SET} operations (string sets, set adds, and list pushes), as these
were part of the default benchmark operations test suite.

Table~\ref{tab:overhead} shows the steady state overhead of this experiment.
We show unmodified \redis in column 3 for comparison and broke the overhead
into separate categories: \tool with no prefixes to update declared (causing \tool to return
immediately for each key) in column 4, \tool with a single prefix declared
(causing a hash lookup and a version check for each key) in column 5, and \tool
with a previous prefix declared but no previous keys with the old prefix 
(causing a hash lookup, a version check, and a string concatenation to look for
a non-existent previous key) in column 6.  Each sub-column of
Table~\ref{tab:overhead} shows the total time for the test, the siqr
(Semi-Interquartile Range) to show the variance, and the overhead as a
comparison against unmodified \redis.  We ran this benchmark many times with various
configurations (multiple key prefixes to track, less or fewer keys, less or fewer clients, etc) and
found that the overhead varied generally around $\pm3\%$, with no consistent pattern
between any of the tests, even repeated tests with the exact same setup.  
The numbers presented in the table show some
negative and some positive overhead, reflecting this variation. Notice that the siqr numbers
show that the variance is relatively high, as high as 1.49s for setting strings
with \tool and a prefix, shown in the fourth row of the fifth column.

The bottom half of Table~\ref{tab:overhead} shows a modification of the original overhead
experiment, using a pipeline to feed 10 instructions into each round trip to
\bench over localhost. This reduced the I/O overhead, putting more emphasis on
\tool operations.  We found that these numbers showed a bit more overhead, and
allowed us to bound the overhead at $5.74\%$ for 10 subsequent pipelined
instructions.  This test demonstrated that although there is some overhead
added by \tool, for the non-pipelined version and most commonly-used scenario
(Table~\ref{tab:overhead}), the overhead is mostly buried in I/O and very low
overall.  (In our test programs, described next, Amico pipelined at most 3
instructions per round trip, and \redisFS did not use pipelineing.) 

In addition to time overhead, \tool incurs some additional memory overhead due
to tracking the version information.  Table~\ref{tab:rss} shows the maximum
resident set size as reported by \emph{ps}.  Empty, \redis and \tool take up
about the same amount of size in memory.  With 1 million keys each mapping to
10-byte values and with 5 separate prefixes declared, \tool takes up about 16.5MB
(\tid 15\%) more memory than unmodified \redis.  This includes the extra version
field (4 bytes) on each value structure, the amount of space it takes to store
the version lookup information and hash table, and any extra padding that may
be automatically added to the additional structures.  

\subsection{Redisfs}\label{sec:redisfs}
\begin{figure}[t]
  \begin{centering}
  \includegraphics[scale=0.60]{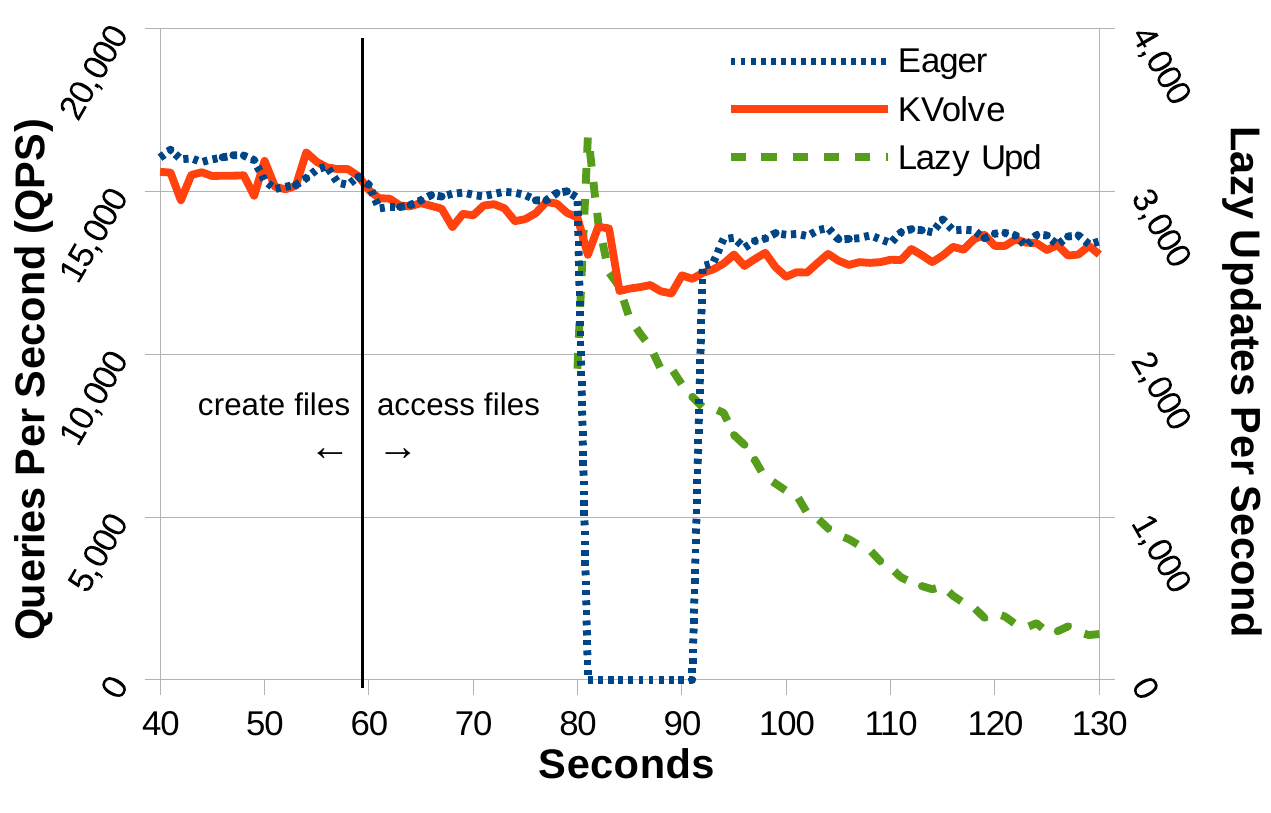}
  \vspace{-1mm}
  \caption{Lazy vs. eager updates for RedisFS}
  \label{graph:redisfs}
  \end{centering}
  \vspace{-5mm}
\end{figure}

Redisfs~\cite{redisfs} uses \redis as the backend to the FUSE~\cite{fuse} file
system.  The inode information, directory information, and all file system data
are stored in \redis. On startup, FUSE mounts a directory with \redis as the
backend, and a user can perform all of the normal operations of a file system,
with the data silently being stored in \redis. Redisfs has 8 releases, \tid2.2K
lines of C code each. In \redisFS .5, released March 4th, 2011, file data is
stored in a \redis as a binary string with no compression, and the directory
keys have the format \code{skx:/path/todir}.  In \redisFS .7, released March
11th, 2011, file data is compressed using \code{zlib}, and directory keys have
the format \code{skx:DIR:/path/todir}.  (Note that \redisFS .6 contained an
error and was retracted, so we use versions .5 and .7.)  This change makes it
impossible to view the directories or any of the files using \redisFS .7 for
any files created using \redisFS .5.


In all versions, the inode data is stored across about 12 \redis
keys with meta information such as modification time, file size,
and also the data itself.  All file system information is represented in \redisFS with
four prefixes: the \code{skx:/} prefix for directories (which
is updated to \code{skx:DIR/} in \redisFS .7), the \code{skx:NODE} prefix
for inodes (some of which is updated to add compression in \redisFS
.7), \code{skx:PATH} for paths to directories, and \code{skx:GLOBAL} to
track internal structure; the last two are not updated.
To make \redisFS compatible with \tool, we added only 6 lines of code in
both versions which consisted of an additional call to \redis on start-up
to declare that we would be using those 4 prefixes at either version .5 or .7,
along with a few additional lines of error handling to make sure that the prefixes 
were properly set.  

We performed an update from \redisFS .5 to \redisFS .7, both by migrating the
keys offline (referred to as the Eager version), and with \tool to
automatically rename the directory keys as they are accessed and to add
compression to the files as they are accessed. In addition to updating \redisFS
with \tool, we also used Kitsune~\cite{DBLP:journals/toplas/HaydenSSHF14},
whole-program update software for C, to allow us to also dynamically update
\redisFS along with the data so that the users experience no downtime; the
switchover from .5 to .7 is completely seamless.  Normally, killing
\redisFS .5 and restarting at \redisFS .7 also causes the mount point to be
unmounted then remounted (causing the user to have to switch back into the
mounted directory after remount), but with Kitsune, the mount point is not
disrupted during the switchover.  We used the file system benchmark
PostMark~\cite{postmark} to generate a workload for \redisFS, creating an
initial 10,000 files ranging from 4-1024 bytes in 250 subdirectories plus the
root directory, for a total of 251 directories.  We ran PostMark outside the
root directory mount point, accessing the files via full path name to avoid having
to change directories due to the restart for the Eager (non-\tool/Kitsune) version. 

Figure~\ref{graph:redisfs} shows the results of the \redisFS experiment. After
about 60 seconds, PostMark switched from creating the new files to reading from
or appending to existing files.  As shown on the \emph{left} y-axis, both 
\tool and the Eager version had a very
similar average Queries Per Seconds (QPS), displayed by the solid and finely
dashed lines.  At 80 seconds, we killed
\redisFS .5.  For \tool, we used Kitsune to dynamically update to \redisFS .7
without pause, maintaining the mount point so that the benchmark never lost
access to the files or the directory structure, and \tool continued to process
queries throughout the update. 
For the Eager version, we halted all traffic to \redis and migrated the data,
performing the renames and compression as necessary. In this update, not all
of the keys needed to be updated, only the 251 directory keys that needed to be
renamed and the 10,000 data keys that needed to be compressed. However, the database
contained 123,002 total keys, and the to-be-updated keys were searched for in the
database, adding to the pause time.  This offline update process took about 12
seconds, as shown in Table~\ref{tab:pausetime}. 

In addition to showing the QPS lines, the green widely-dashed line in
Figure~\ref{graph:redisfs} shows the number of lazy updates per second for
\tool, corresponding to the \emph{right} y-axis. Immediately after the update, this
number burst to $\sim$3K keys per second, and quickly trailed off as keys
were lazily updated. \tool renamed the 251 directory keys, updated the version on all 112,752 keys in
the \code{skx:INODE} prefix, and compressed the data for the 10K keys in that prefix that
contained file data. 

Overall the impact on the update experienced by \redisFS was minimal, as the
QPS dipped only slightly right after the update before it quickly returned to
full speed around the 120 second mark of the experiment.  After the update, the
overall QPS was slower for both \tool and \redisFS because the files must be
compressed and decompressed as they were accessed.

\subsection{Amico}\label{sec:amico}
\begin{figure}[t]
  \begin{centering}
  \includegraphics[scale=0.60]{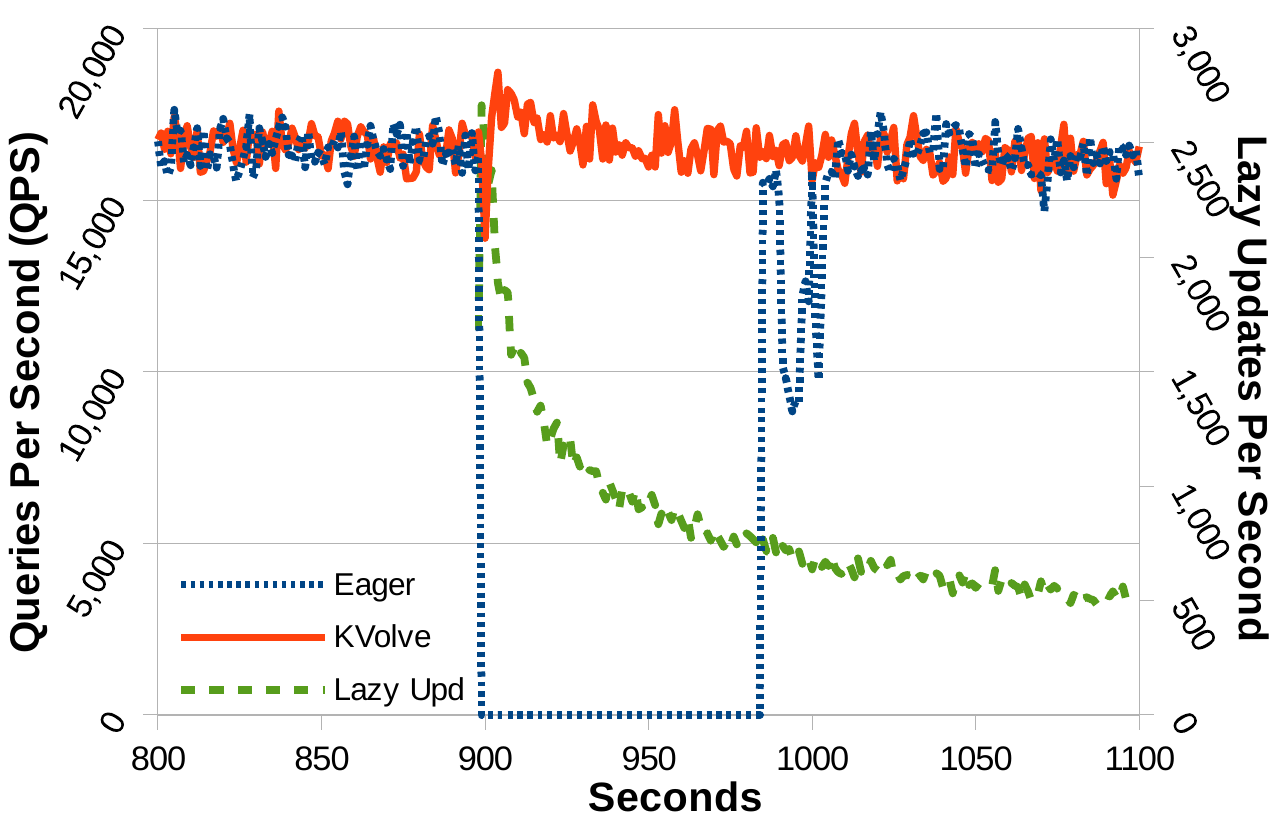}
  \vspace{-1mm}
  \caption{Lazy vs. eager updates for Amico}
  \label{graph:amico}
  \end{centering}
  \vspace{-5mm}
\end{figure}

Amico~\cite{Amico} maps relationships in the style of a social network,
defining a set of users and the relationships between them.  Amico provides an
API that allows queries over a data set of users: a user may be following or be
followed by any number of other users.  Amico is backed by \redis, has 10
versions created between 2012-2013, and is written in \tid200 lines of Ruby code.
Amico version 1.2.0, released Feb 22, 2012, stores these relationships in 5
different types of \redis keys with the following prefixes:
\code{amico:followers}, \code{amico:following}, \code{amico:blocked},
\code{amico:reciprocated}, and \code{amico:pending}.  In version 2.0.0,
released Feb 28, 2012 (the next consecutive version after 1.2.0), the
developers added the concept of a ``scope'' so that there could be different
graphs stored in \redis with prefixes to keep them separate, such as ``school''
network and a ``home'' network. The default name for the scope is ``default'',
such that all of the keys are prefixed with \code{amico:followers:default} for
example.  This change makes databases created with Amico 1.2.0 incompatible
with Amico 2.0.0. To make Amico work with \tool, we only changed the same 4
lines of code in each version to declare the prefixes right after Amico
connects to \redis.

For this experiment, we used the LiveJournal data set from the
SNAP~\cite{snapnets} library.  The LiveJournal data set has 4,847,571 nodes and
68,993,773 directed edges defined by ordered node id numbers \emph{A follows B}
such as \code{186032 2345471}, which we shuffled into two separate files for
reading in a random order. To create a workload, we started two programs with
calls to Amico 1.2.0: one program to read from the first random
file and add nodes to the Amico network, and one program to read from the
second random file and perform queries over nodes in the network such as
querying if \code{USER A} followed \code{USER B} or querying the number of
followers of \code{USER A}.  After letting the programs run for 900 seconds (15
minutes), the \redis database was filed with 792,711 keys containing nodes and
edge data.

At the 900 second mark, as shown in Figure~\ref{graph:amico}, we stopped both
of the Amico 1.2.0 programs.  For the Eager case (finely dashed line), we then
updated all 792,711 keys, renaming them to have the \code{default} scope prefix
in all of the key names. This migration took $\sim$87 seconds as shown in
Table~\ref{tab:pausetime}. In addition to the pause, the Eager case shows a
continued disruption until around the 1,000 second mark.  After the migration
was complete, we started the same writer/reader programs, this time using
Amico 2.0.0.  For the \tool case (solid line), we immediately started the
two Amico 2.0.0 programs after the update so that the keys could be lazily
migrated. Right at the update point, there is a \tid2K drop in the QPS (\emph{left} y-axis),
before a brief spike and a return to the original rate.  The widely-dashed green line
corresponds to the \emph{right} y-axis and shows the number of lazy updates that take
place each second.  Because this is a very large data set, many of the keys are
not accessed immediately, taking full advantage of laziness.  Although the lazy
updates continue at a rate of about 500 per second at the 1,100 second mark, this does
not significantly impact overall queries per second, as shown by the solid line
maintaining a similar QPS before and after the update.

\begin{table}[]
\centering
\caption{Offline update pause times}
\begin{tabular}{|l|l|l|}
\hline
         & \textbf{Pause (s)} & \textbf{Update Events}     \\ \hline
Amico    & 87s         & 792,711 : rename                  \\ \hline
\redisFS & 12s         & 10,000 : compress, 251 : rename   \\ 
         &             & (123,002 total keys in database)  \\ \hline
\end{tabular}
\label{tab:pausetime}
\vspace{-4 mm}
\end{table}

\section{Related Work}
\label{sec:related}




In the realm of relational databases, the evolution of an application's schema is characterized by the 
changes to the {\tt CREATE TABLE} statements used to instantiate the schema in subsequent versions of the application. 
%
%
In practice, complex schema changes often require taking the application offline or locking the database tables, 
such as the update to Wikipedia that held a write lock for 22 hours~\cite{Wikipedia2005Downtime}. 
%
%
%
Prior research has proposed supporting non-blocking schema changes by
accepting out of date copies of database objects~\cite{DBLP:journals/vldb/StonebrakerALPSSSY96}
or by implementing changes on-the-fly using 
triggers~\cite{DBLP:conf/icde/Ronstrom00}
or log redo~\cite{DBLP:conf/edbt/LolandH06}.
Additionally, several professional tools can perform {\tt ALTER TABLE} operations in a non-blocking manner \cite{openarkkit, facebook, pt-online, LHM, tablem}.
%
Because these tools focus only on the database, the changes implemented must be backward compatible to avoid breaking the application logic. 
To avoid this limitation, the Imago system~\cite{Dumitras:2009} proposed installing the new version in a parallel universe, with dedicated application servers and databases, which allowed it to perform an end-to-end upgrade atomically.
This can be achieved in practice by deploying parallel AppEngine~\cite{appengine} applications, at multiple versions. 
However, this approach duplicates resources and exposes the new version to the live workload only after the data migration was completed.

In contrast, the F1 database from Google implemented an asynchronous protocol~\cite{F1} for adding and removing tables, columns and indexes, which allows the servers in a distributed database system to access and update all the data during a schema change and to transition to the new schema at different times. 
%
%
This is achieved 
by having stateless database servers with temporal schema leases, 
by identifying which schema-change operations may cause inconsistencies, 
and by breaking these into a sequence of schema changes that preserve database consistency as long as servers are no more than one schema version behind. 
%
Google's Spanner distributed key-value store~\cite{corbett2013spanner} (which provides F1's backend) supports changes to key formats and values by registering schema-change transactions at a specific time in the future and by utilizing globally synchronized clocks to coordinate reads and writes with these transactions. 
These systems do not address 
changes to the format of Protobufs stored in the F1 columns or Spanner values~\cite{sadalage} or 
inconsistencies that may be caused by interactions with (stateful) clients using different schemas~\cite{Tudor10:UpgradeOrNot}. 



Schema evolution in NoSQL databases is less well understood, as these databases do not provide a data definition language for specifying the schema. 
However, many applications attach meaning to the format of the keys and values stored in the database, and these formats may evolve over time. 
In particular, the values often correspond to data structures serialized using JSON~\cite{json} or a binary format like Thrift~\cite{thrift}, Protobufs~\cite{protobuf}, or Avro~\cite{avro}.
The latter formats 
have schema-aware parsers, which include 
some support for schema changes, 
e.g. by skipping unknown fields or by attempting to translate data from the writer schema into the reader schema~\cite{kleppmann}.
However, orchestrating the actual changes to the data and the application logic is entirely up to the programmer.
%
%

One approach to defining schema changes defines a declarative schema evolution language for NoSQL data\-bases~\cite{Scherzinger}. 
%
%
This language allows specifying more comprehensive schema changes and enables the automatic generation of database queries for migrating eagerly to the new schema.
(While the paper also mentions the possibility of performing the migration in a lazy manner, which is needed for avoiding downtime, design and implementation details are not provided.) 
Other approaches use a domain-specific language (DSL) for describing data schema migrations for Python~\cite{saur15morpheus} and for Haskell datatypes~\cite{gundry}. 
Many other approaches \cite{Rahm11,Curino,Velegrakis,An:2008} have focused on the problem of synthesizing the transformation code to migrate from one
schema version to the next, and the transformation is then typically applied offline, rather than incrementally online. 
In this paper, we 
focus on how to
apply a transformation without halting service rather than synthesizing the transformation code.

In practice, developers are often advised to 
handle all the necessary schema changes in custom code, added to the application logic that may modify the data in
the database~\cite{sadalage, servicestack, rethans, stack_ov}.
%
This approach burdens programmers with  
complex code that mixes application and schema-maintenance logic and 
does not provide a mechanism for reasoning about the correctness of schema changes performed concurrently with the live workload.

Our work is also related to the body of research on dynamic software updates~\cite{DBLP:journals/toplas/HaydenSSHF14,DBLP:conf/oopsla/PinaVH14,DBLP:conf/middleware/GiuffridaIT14,Chen:2007:PPL:1248820.1248860},
which aim to modify a running program on-the-fly, without causing downtime. 
However, with the exception of a position paper~\cite{despande05dbupdate},
these approaches focus on changes to code and data structures loaded in memory, 
rather than changes to the formats of persistent data stored in a database. 
%



\section{Conclusions and Future Work}

This paper has presented \tool, a general approach to evolving a NoSQL database
without downtime. \tool adapts \redis to migrate data as it is accessed,
reducing downtime that would otherwise result during a data upgrade, and
minimizing required changes to applications. We find that \tool imposes
essentially no overhead when not performing an
update, and minimal overhead when performing an update. 

In the future, we would like to expand \tool to work with \redis Cluster, a
distributed implementation of \redis.  We also would like to add direct support for programmer-specified,
backward-compatible updates, which would support continued operation without
restarting clients. Finally, we would like to streamline writing the
transformation function with a DSL, simplifying the
update planning process. 

We plan to release our code and make it freely available.

\bibliographystyle{./IEEEtran}
\bibliography{bib}

\end{document}